\documentclass[preprint,preprintnumbers,amsmath,amssymb,showkeys]{revtex4}
\usepackage{dcolumn}% Align table columns on decimal point
\usepackage{bm}% bold math
\usepackage{epsfig,subfigure}

\begin{document}

\preprint{}

\title{Insights and possible resolution to the information loss paradox via the tunneling picture}

\author{Douglas Singleton}
\email{dougs@csufresno.edu}
\affiliation{Physics Department, CSU Fresno, Fresno, CA 93740-8031, USA}

\author{Elias C. Vagenas}
\email{evagenas@academyofathens.gr}
\affiliation{Research Center for Astronomy and Applied Mathematics,
Academy of Athens,\\ Soranou Efessiou 4, GR-11527, Athens, Greece}

\author{Tao Zhu}
\email{zhut05@lzu.cn}
\author{Ji-Rong Ren}
\email{renjr@lzu.edu.cn}
\affiliation{Institute of Theoretical
Physics, Lanzhou University, Lanzhou 730000, P. R. China}

%\date{\today}

\begin{abstract}
This paper investigates the information loss paradox in the WKB/tunneling
picture of Hawking radiation. In the tunneling picture one can obtain
the tunneling amplitude to all orders in $\hbar$. However all terms
beyond the lowest, semi-classical term involve
unknown constants. Despite this we find that one can still arrive at
interesting restrictions on Hawking radiation to all orders in $\hbar$:
(i) Taking into account only quantum corrections the spectrum remains thermal
to all orders. Thus quantum corrections by themselves will not resolve the
information loss paradox. (ii) The first quantum correction give a
temperature for the radiation which goes to zero as the mass of the black hole goes to zero.
Including higher order corrections changes this nice result of the first order corrections.
(iii) Finally we show that by taking both quantum corrections and back reaction into account it is possible
under specific conditions to solve the information paradox by having the black hole evaporate completely
with the information carried away by the correlations of the outgoing radiation.
\end{abstract}


\maketitle

\section{Introduction}

There are many derivations of Hawking radiation in literature.
Recently this phenomenon has been studied using a variant of
the WKB/tunneling approach \cite{Wilczek}. This approach provides a simple and physically
intuitive picture. The tunneling method has two
versions: (i) the null geodesic method \cite{Wilczek}; (ii) the
Hamilton-Jacobi method \cite{padma}. The original work on the tunneling method
was unable to obtain the thermal nature of the Hawking radiation. This shortcoming
was recently solved by Banerjee and Majhi \cite{blackbody} who obtained the thermal
spectrum for Hawking radiation using the tunneling picture. The thermal nature of
Hawking radiation leads to the information loss
paradox: since the radiation is thermal there are no correlations between the
emitted field quanta and so one losses information about the nature of the matter that originally
formed the black hole. More technically, the complete evaporation of a black hole, whereby a pure
quantum state evolves into a thermal state, would violate the quantum mechanical unitarity.
A recent review of the information paradox
can be found in reference \cite{information paradox}.

There have been various proposals
for resolving this information loss paradox which usually involve the idea that Hawking radiation
is modified when one goes beyond the approximations used in the various methods of deriving it. One possibility
is that higher order corrections in $\hbar$ may make the spectrum non-thermal. For example, in the case
of charged particle pair creation in a uniform electric field -- the Schwinger effect \cite{schwinger} which can be thought of
as a non-GR, field theory version of Unruh/Hawking radiation \cite{holstein} -- the tunneling amplitude can
be calculated exactly to all orders in $\hbar$ and it is non-thermal. Here we show that the same
is not true for Hawking radiation -- using the tunneling formalism it can be shown that to all
orders of $\hbar$ the radiation spectrum is thermal. Thus, even taking the full quantum theory into account
does not resolve the information paradox \footnote{By full quantum theory we mean the corrections to all orders in
$\hbar$ coming from the non-gravitational fields. The metric we consider is still completely classical so that gravity is
not quantized.}.

Quantum tunneling with back reaction gives the relation between the tunneling rate and entropy change of black hole
\cite{Wilczek}. This relation provides an alternative approach to the information
loss paradox \cite{vagenas,information}. However, the considerations are confined to the semi-classical
approximation. The approach to the information loss paradox given here involves both quantum corrections to
all orders and back reaction. For this approach we look at the information loss paradox
in terms of entropy. The original black hole has some entropy
$S$ which is proportional to the surface area, $A$, of the black hole \cite{bekenstein}. If the
black hole evaporates away completely into thermal radiation then the total entropy after the evaporation
will be greater than that before. From the microscopic view of entropy this indicates that the number of
microstates is greater after evaporation as compared to
before evaporation. This increase in the number of states indicates
that the evolution is non-unitary. In order to have unitary evolution the entropy before
should be equal to the entropy after. In this article we will show that it is possible
under specific conditions for the entropy
before, as determined by the surface area of the black hole, to be made equal to the entropy of
the outgoing radiation after the black hole has completely evaporated. The entropy of the outgoing
radiation will be calculated from the non-trivial correlations between emitted field quanta which
arise due to higher order quantum corrections plus back reaction.

\section{Blackbody spectrum beyond the semi--classical approximation}

In this section we briefly review the salient results from references
\cite{beyond, beyond1} where a formalism was developed to investigate
quantum corrections {\it to all orders} in $\hbar$ in the tunneling method.

We begin by considering a general class of static, spherically symmetric space--time
\begin{eqnarray}\label{spherically symmetric metric}
ds^2=-f(r)dt^2+\frac{dr^2}{f(r)}+r^2d\Omega^2.
\end{eqnarray}
The horizon of the black hole $r=r_H$ is given by $f(r_H)=0$. In
this space--time, a massless scalar particle obeys the Klein-Gordon
equation
\begin{eqnarray}\label{Klein-Goedon equation}
-\frac{\hbar^2}{\sqrt{-g}}\partial_\mu(g^{\mu\nu}\sqrt{-g}\partial_\nu)\phi=0.
\end{eqnarray}
In the tunneling framework, the tunneling particle is
considered as a spherical shell so that the trajectory of the
tunneling process is radial. Therefore only the $(r,t)$ sector of
the metric (\ref{spherically symmetric metric}) is important and
thus tunneling of a particle from a black hole can be
considered as a two-dimensional quantum process in the $(r,t)$ plane.
For a two dimensional theory, the standard WKB ansatz for the wave
function $\phi$ can be expressed as
\begin{eqnarray}\label{scalar field WKB}
\phi(r,t)=\exp\left[\frac{i}{\hbar}I(r,t)\right]
\end{eqnarray}
where $I(r,t)$ is the one particle action which can be expanded in
powers of $\hbar$ as
\begin{eqnarray}\label{expansion of the action}
I(r,t)=I_0(r,t)+\sum_j \hbar^j I_j(r,t).
\end{eqnarray}
Here $I_0(r,t)$ is the semi--classical action and the others are
quantum corrections. As shown in \cite{beyond,beyond1}, $I_i(r,t)$ are proportional to
$I_0(r,t)$, so that (\ref{expansion of the action}) can be written as
\begin{eqnarray}\label{expansion-elias}
I(r,t)=\left(1+\sum_j\gamma_j \hbar^j \right)I_0(r,t)
\end{eqnarray}
with $\gamma _j$ being unknown proportionality constants. We are using units with
$G=c=k_B =1$ so that $\hbar$ has units of length square -- the Planck length
$l_P$. The solution of the semi--classical
action $I_0$ has the form
\begin{eqnarray}
\label{in-out}
I_0(r,t)=\omega (t\pm r_*),~~~~r_*=\int\frac{dr}{f(r)}.
\end{eqnarray}
The $-$ ($+$) sign refers to outgoing (ingoing) trajectories.
For the high order correction terms in equation (\ref{expansion-elias}),
the undetermined proportionality coefficients $\gamma_j$
have dimension $\hbar^{-j}$. For concreteness we consider
a Schwarzschild black hole whose only macroscopic parameter
can be defined in terms of the radius of the horizon $r_H$.
Performing some elementary dimensional analysis -- requiring that $\gamma_j \hbar ^j$
be dimensionless and remembering that in our units $\hbar$ has units of
$l_P ^2$ -- one can show the coefficients $\gamma_j$ have dimension $r_H^{-2j}$.
This is expressed as
\begin{eqnarray}
\gamma_j=\frac{\alpha_j}{r_H^{2j}}
\end{eqnarray}
with $\alpha_j$ dimensionless constants. Thus the solution for the scalar field $\phi(r,t)$ is,
\begin{eqnarray}\label{solution}
\phi(r,t)=\exp{\left[\frac{i}{\hbar} \left( 1+\sum_j\frac{\alpha_j\hbar^j}{r_H^{2j}} \right) \omega
(t\pm r_*)\right]}.
\end{eqnarray}
If $(t\pm r_*)$ has an imaginary part then $\phi (r, t)$ will have an exponentially decreasing
part from which the tunneling amplitude, the temperature and the spectrum of the radiation
can be found. From equation (\ref{in-out}) and performing a contour integration
it has been shown \cite{chow} that
$r_*$ does have an imaginary part but gives twice the correct Hawking
temperature. In \cite{time} it was shown that in general there is also an imaginary contribution coming from
the time part of $(t\pm r_*)$. This is seen by introducing the Kruskal time (T) and space (X) coordinates
inside and outside the horizon
\begin{eqnarray}
T_{\texttt{in}}&=& \sqrt{1- 2\kappa  r} ~ e^{\kappa r}\cosh{(\kappa
t_{\texttt{in}})},~~~~~~~~X_{\texttt{in}}= \sqrt{1- 2 \kappa  r} ~ e^{\kappa r }\sinh{(\kappa t_{\texttt{in}})},\nonumber\\
T_{\texttt{out}}&=& \sqrt{2 \kappa r -1 } ~ e^{\kappa r}\sinh{(\kappa
t_{\texttt{out}})},~~~~~X_{\texttt{out}}= \sqrt{2 \kappa r -1} ~ e^{\kappa r} \cosh{(\kappa
t_{out})},
\end{eqnarray}
where $\kappa=\frac{f'(r_H)}{2}$ is the surface gravity of the black
hole. The inside ($T_{\texttt{in}},X_{\texttt{in}}$) and outside
($T_{\texttt{out}},X_{\texttt{out}}$) coordinates can be connected
by the following transformation
\begin{eqnarray}
t_{\texttt{in}}=t_{\texttt{out}}+\frac{i\pi}{2\kappa}.
\end{eqnarray}
Thus, on crossing the horizon the time part picks up an imaginary part.
This temporal contribution plays an important role in determining the
correct Hawking temperature for black holes in the tunneling
method \cite{time}. Under this transformation the inside and outside
modes of the scalar field (\ref{solution}) are connected by
\begin{equation}\label{transf}
\phi_{\texttt{in}}^{\texttt{R}}= \exp{ \left[-\frac{\pi\omega}{\hbar\kappa} \left (1+\sum_j\frac{\alpha_j\hbar^j}{r_H^{2j}} \right) \right]}
\phi_{\texttt{out}}^{\texttt{R}}~,~~~~~~~
\phi_{\texttt{in}}^{\texttt{L}} = \phi_{\texttt{out}}^{\texttt{L}}.
\end{equation}
Here $\texttt{in,out}$ means modes inside/outside the horizon and $\texttt{R,L}$ indicate modes moving to the right/left
(away/toward $r=0$). We now review the parts of the derivation of the spectrum \cite{blackbody} that will be
important to showing that to all orders in $\hbar$ the spectrum still remains thermal.
Consider the physical state with $n$ non-interacting virtual
pairs created inside the black hole. When observed from outside, this state is given by
\begin{eqnarray}
|\Psi\rangle=N\sum_n|n_{\texttt{in}}^{\texttt{L}}\rangle\otimes|n_{\texttt{in}}^{\texttt{R}}\rangle
=N\sum_n \exp{ \left[-\frac{n \pi\omega}{\hbar\kappa} \left (1+\sum_j\frac{\alpha_j\hbar^j}{r_H^{2j}}
\right) \right]}|n_{\texttt{out}}^{\texttt{L}}\rangle\otimes|n_{\texttt{out}}^{\texttt{R}}\rangle\
\end{eqnarray}
where $N$ is the normalization constant and
$|n^{\texttt{L(R)}}\rangle$ represents $n$ number of the left(right)-moving mode. In the above expression, we have used the relations
(\ref{transf}). $N$ can be determined from the normalization
condition, which leads to
\begin{equation}
N= \sqrt{1 \mp \exp \left[ {-\frac{2\pi\omega}{\hbar\kappa} \left( 1+\sum_j\frac{\alpha_j\hbar^j}{r_H^{2j}} \right)} \right] }
\end{equation}
where the $-$ ($+$) sign is for bosons (fermions). One can define the density matrix operator for
the system as
\begin{eqnarray}
&~& \rho_{\texttt{boson, fermion}} = |\Psi\rangle_{(\texttt{boson,fermion})}\langle\Psi|_{(\texttt{boson,fermion})}\nonumber\\
&=& \left[ 1 \mp e^{-\frac{2\pi\omega}{\hbar\kappa} \left( 1+\sum_j\frac{\alpha_j\hbar^j}{r_H^{2j}} \right) } \right]
\sum_{m,n}e^{-\frac{(n+m)\pi\omega}{\hbar\kappa} \left( 1+\sum_j\frac{\alpha_j\hbar^j}{r_H^{2j}} \right) }
|n_{\texttt{out}}^{\texttt{L}}\rangle\otimes|n_{\texttt{out}}^{\texttt{R}}\rangle\langle
m_{\texttt{out}}^{\texttt{R}}|\otimes\langle
m_{\texttt{out}}^{\texttt{L}}|.
\end{eqnarray}
It should be stressed that the left-moving modes are going in the direction of $r=0$ and thus
they are not observed by an outside observer. Therefore, one can trace out the left-moving modes,
giving the density matrix operator for the right-moving modes as
\begin{equation}
\rho^{\texttt{R}}_{\texttt{boson,fermion}}= \left( 1 \mp e^{-\frac{2\pi\omega}{\hbar\kappa}
\left( 1+\sum_j\frac{\alpha_j\hbar^j}{r_H^{2j}} \right) } \right)
\sum_{n}e^{-\frac{2n\pi\omega}{\hbar\kappa} \left( 1+\sum_j\frac{\alpha_j\hbar^j}{r_H^{2j}} \right) }|n_{\texttt{out}}^{\texttt{R}}\rangle\langle
n_{\texttt{out}}^{\texttt{R}}|.
\end{equation}
From this one can immediately get the average number of
particles detected at asymptotic infinity
\begin{equation}
\label{n-spectrum}
\langle
n\rangle_{(\texttt{boson,fermion})}=\texttt{trace} \left( \hat{n}\hat{\rho}^{\texttt{R}}_{(\texttt{boson,fermion})} \right)
=\frac{1}{e^{\frac{2\pi\omega}{\hbar\kappa} \left( 1+\sum_j\frac{\alpha_j\hbar^j}{r_H^{2j}} \right)} \mp 1}.
\end{equation}
This result is the Bose-Einstein distribution (for $-$ sign) or Fermi-Dirac distribution (for $+$ sign).
Both these distributions correspond to a blackbody spectrum with a
corrected Hawking temperature given by
\begin{eqnarray}
\label{corrected Hawking temperature}
T=\frac{\hbar\kappa}{2\pi}\left(1+\sum_j\frac{\alpha_j\hbar^j}{r_H^{2j}}\right)^{-1}.
\end{eqnarray}
In this expression $\frac{\hbar\kappa}{2\pi}$ is
the semi--classical Hawking temperature, $T_H$, and the other terms are
higher order quantum corrections.

We now give an alternative derivation of the quantum corrected temperature in
terms of the tunneling rate, $\Gamma$, since this will be important
for the next section when we make a connection between $\Gamma$ and the
change in entropy. The tunneling rate can be written as
$\Gamma=\frac{P^{\texttt{R}}}{P^{\texttt{L}}}$, where $P^{\texttt{R,L}}$
are the probabilities of the right/left-moving modes to cross the horizon.
From equation (\ref{transf}) one finds
\begin{equation}
P^{\texttt{R}} = | \phi ^{\texttt{R}} _{\texttt{in}} |^2 =
\left| \exp \left[-\frac{\pi \omega}{\hbar \kappa} \left(1 + \sum_j \frac{\alpha _j \hbar ^j}{r^{2j}_H} \right) \right]
\phi ^{\texttt{R}} _{\texttt{out}} \right| ^2 =
 \exp \left[-\frac{2\pi \omega}{\hbar \kappa} \left(1 + \sum_j \frac{\alpha _j \hbar ^j}{r^{2j}_H} \right) \right]~,
\end{equation}
and
\begin{equation}
P^{\texttt{L}} = | \phi ^{\texttt{L}} _{\texttt{in}} |^2 = | \phi ^{\texttt{L}} _{\texttt{out}} |^2 =1~.
\end{equation}
This result for $P^{\texttt{L}}$ shows that with probability $1$ left-moving modes cross the horizon, while
the result for $P^{\texttt{R}}$ gives the probability that modes tunneling out from behind the horizon. Combining
these results and equating the tunneling rate, $\Gamma$, with the Boltzmann distribution, i.e. $e^{\frac{\omega}{T}}$,
\begin{eqnarray}\label{gamma}
\Gamma=\frac{P^{\texttt{R}}}{P^{\texttt{L}}} =
\exp \left[-\frac{2\pi \omega}{\hbar \kappa} \left(1 + \sum_j \frac{\alpha _j \hbar ^j}{r^{2j}_H} \right) \right] =
e^{-\frac{\omega}{T}}~,
\end{eqnarray}
one can again recover the result for the corrected temperature (\ref{corrected Hawking temperature}).

Although the coefficients, $\alpha _j$, are undetermined there are still two important
conclusions one can draw from the results in equations (\ref{n-spectrum}) and
(\ref{corrected Hawking temperature}): (i) The spectrum is still thermal to all orders of $\hbar$.
Thus, the information loss paradox is not resolved by higher order corrections
in $\hbar$ to any order. Note the difference with the Schwinger effect were it is possible to
calculate the decay amplitude to all orders and one finds there that the spectrum for
charge creation in a strong electric field is non-thermal. (ii) In the lowest order one finds that the temperature diverges
as the black hole evaporates, i.e. $T \rightarrow \infty$ as $M \rightarrow 0$. Taking into account the
quantum corrections (and using $\kappa = \frac{1}{4 M}$ and $r_H =2 M$) the temperature in equation (\ref{corrected Hawking temperature}) becomes
$$
T = \frac{\hbar}{8 \pi \left( M + \frac{\alpha _1 \hbar}{4M} + \frac{\alpha _2 \hbar ^2}{16M^3}+...\right) }.
$$
Taking into account the correction to the temperature up to the first quantum correction (i.e.
$\frac{\alpha _1 \hbar}{4M}$ term) one sees that the denominator goes to $\infty$ as
$M \rightarrow 0$, so now $T \rightarrow 0$ as $M \rightarrow 0$. Thus the generic first order quantum correction
appears to modify the lowest order divergence of the temperature as $M \rightarrow 0$. (In the case when $\alpha _1 <0$
there is still some finite $M$ -- i.e. $M= \sqrt{-\alpha _1 \hbar /4}$ -- where $T$ diverges as can be seen from
figure \ref{tb}). Including higher order quantum corrections would appear to also always lead to a zero temperature
as $M \rightarrow 0$. However there are cases -- such as if the $\alpha _i$'s alternate in sign -- when this
result of a zero temperature in the limit $M \rightarrow 0$ is not always the result. In the next section
when we discuss the entropy we will find that in order to have a well behaved entropy as $M \rightarrow 0$
we need to have $\alpha _i$'s which alternate in sign. Thus the apparent nice behavior of the temperature
as $M \rightarrow 0$ is not guaranteed. This point will be discussed further in the next section.

\section{Back reaction and entropy conservation}

In the last section we found that quantum corrections to any order do not resolve the
information paradox but they do tend to make the temperature go to zero, rather than
diverging, as the mass of the black hole is evaporated away to zero.
We now turn to back reaction effects. The tunneling approach
allows one to take into account the effect of the emission of the Hawking
radiation of the horizon in a simple manner using energy conservation \cite{Wilczek}.
Here we combine the back reaction effects along with the $\hbar$
corrections from the previous section to show that it is {\it possible}
under specific conditions to resolve the information paradox.

\begin{figure}[htb]
\begin{center}
\subfigure[$\alpha_1=-\frac{1}{2}$]{\label{A}\includegraphics[width=8cm]{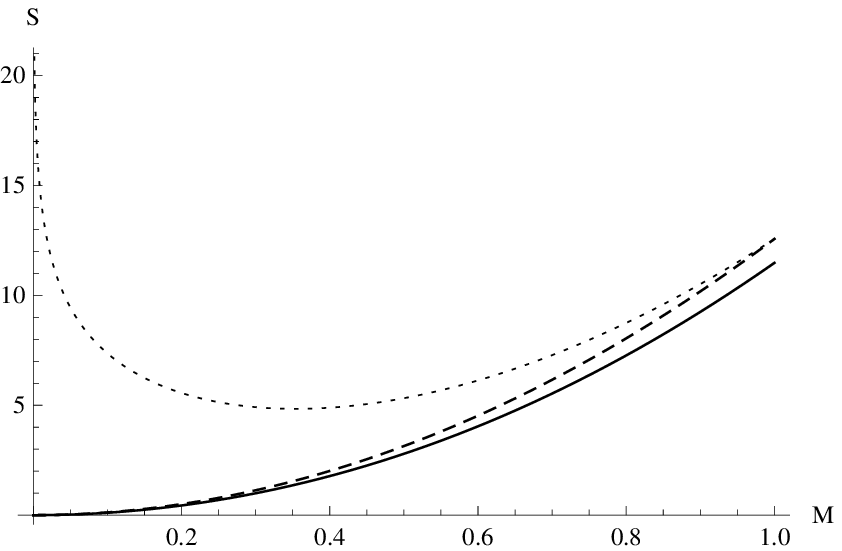}}
\subfigure[$\alpha_1=\frac{1}{2}$]{\label{B}\includegraphics[width=8cm]{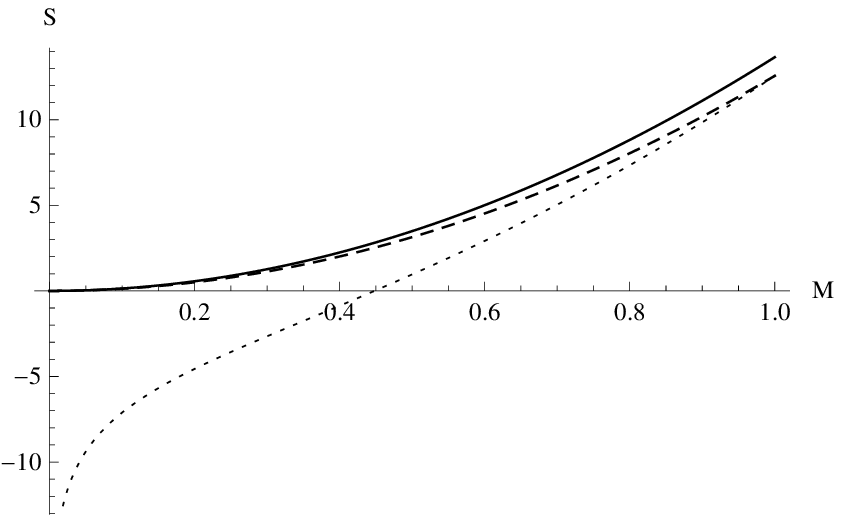}}
\end{center}
\caption{The entropy $S$ versus the black hole mass $M$. The three curves correspond to the lowest order
entropy (dashed curve), the entropy to the first order correction (dotted curve), and the entropy with quantum corrections
to all orders (solid curve) respectively.}
 \label{entropy}
\end{figure}

\begin{figure}[htb]
\begin{center}
\subfigure[$\alpha_1=-\frac{1}{2}$]{\label{ta}\includegraphics[width=8cm]{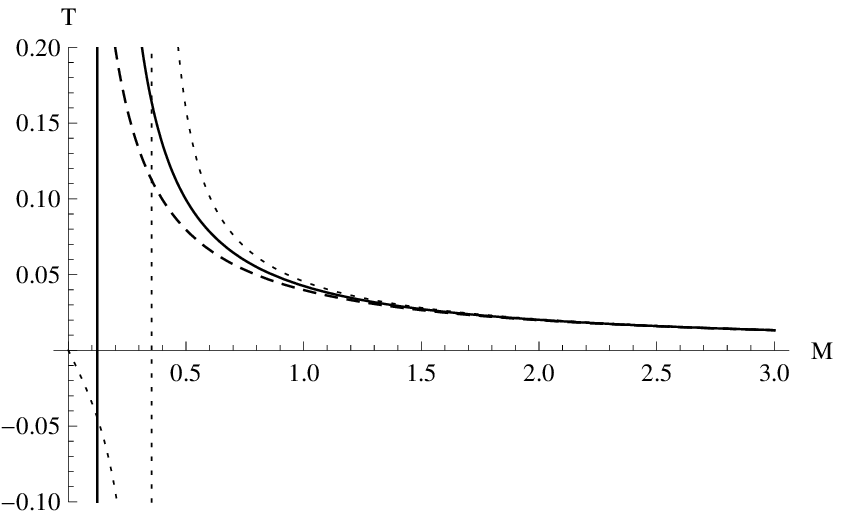}}
\subfigure[$\alpha_1=\frac{1}{2}$]{\label{tb}\includegraphics[width=8cm]{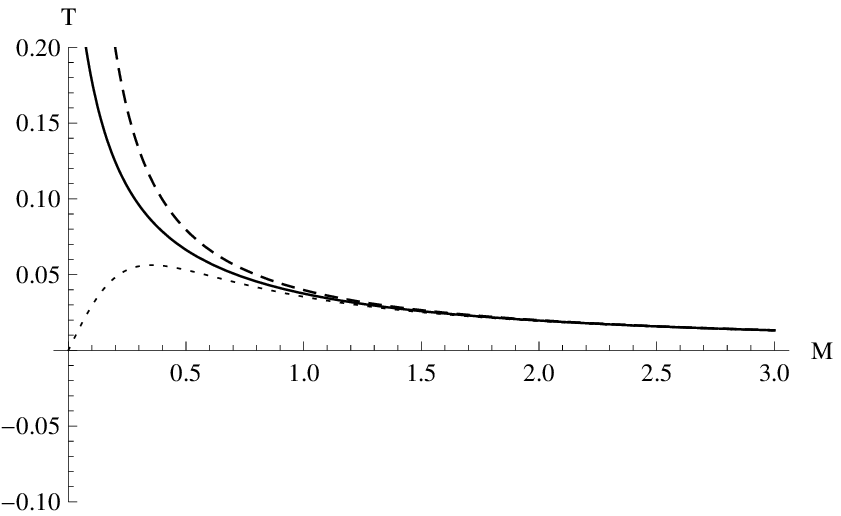}}
\end{center}
\caption{The Hawking temperature $T$ versus the black hole mass $M$. The three curves are corresponding to the lowest order
temperature (dashed curve), the temperature to the first order correction (dotted curve), and the temperature
with quantum corrections to all orders (solid curve) respectively.}
\label{temp}
\end{figure}

One can find an expression to all orders in $\hbar$ for the entropy by integrating the first law of thermodynamics, namely
$dM = T dS$ where $M$ is the initial mass of the black hole, $T$ is temperature and
$S$ is entropy. Thus, $S = \int \frac{dM}{T}$ which upon substituting the expression for the
modified temperature from equation (\ref{corrected Hawking temperature}) gives the modified entropy as a function of $M$
\begin{equation}
\label{entropy-corrected}
S (M)  = \frac{4 \pi}{\hbar} M^2 + \pi \alpha _1 \ln \left( \frac{M^2}{\hbar} \right)
- \pi \sum _{j=1} ^\infty \frac{ \alpha_{j+1}}{4^j j} \left( \frac{\hbar}{M^2} \right) ^j.
\end{equation}
Now in general as $M \rightarrow 0$ this corrected expression for $S$ diverges unless
all $\alpha _j =0$ in which case $S\rightarrow 0$. This has led to the idea that the black hole
evaporation will stop with a black hole ``remnant" of some undetermined mass $m_R$ \cite{remnant}. Here we would like to
point out another possibility. We impose the condition that the unknown $\alpha_j$'s take values
that allow $S \rightarrow 0$ as $M \rightarrow 0$. This condition can be met by setting
\begin{equation}
\label{alpha}
\alpha _{j+1} = \alpha _1 (-4)^j ~~ \text{for} ~~ j=1,2,3... ~~.
\end{equation}
Employing this condition, the sum in equation (\ref{entropy-corrected}), i.e. the third term,
becomes equal to $+\alpha _1 \pi \ln (1 + \hbar/M^2)$ and the entropy now reads
\begin{equation}
\label{entropy-corrected2}
S (M)  = \frac{4 \pi}{\hbar} M^2 + \pi \alpha _1 \ln \left( 1 + \frac{M^2}{\hbar} \right)~.
\end{equation}
It is easily seen that for $M \rightarrow 0$ the entropy tends to zero, i.e. $S \rightarrow 0$.
The behavior of entropy $S(M)$ versus black hole mass $M$
have been shown in figure \ref{entropy}. Strictly the identification of the sum in equation (\ref{entropy-corrected})
with $\alpha _1 \pi \ln (1 + \hbar/ M^2)$ is only valid for $\sqrt{\hbar} < M$, i.e. when the
mass, $M$, is larger than the Planck mass. However we can use analytic continuation to define the sum via
$\alpha _1 \pi \ln (1 + \hbar/M^2)$ even for $\sqrt{\hbar} > M$. This is analogous to the
trick in String Theory \cite{zwiebach} where the sum $\sum _{j=1} ^\infty j$ is defined as $\zeta (-1) = -1/12$
using analytic continuation of the zeta function $\zeta (s) = \sum _{n=1} ^\infty n^{-s}$. Although the choice
of $\alpha_j$'s in equation (\ref{alpha}) is done by hand, one can view this in the following way: When a consistent theory
of quantum gravity is found it should give the full expression for $S(M)$. For example, Loop Quantum Gravity
gives $\alpha _1 = - 1/2$ \cite{meissner}. By setting the physical condition that $S (M=0) \rightarrow 0$
we find what the $\alpha _j$'s from a complete theory of quantum gravity should be. This is in some sense
``looking at the back of the book for the answer". One can ask how unique is the choice in equation (\ref{alpha})? Are there
other choices which would yield $S (M=0) \rightarrow 0$? As far as we have been able to determine there are no other choices
of $\alpha _j$'s that would give $S (M=0) \rightarrow 0$, but for the time being we can not provide a formal proof.
In regard to the temperature the choice of $\alpha _j$'s in equation (\ref{alpha}) leads to a geometric series
and $T$ takes on the following form
\begin{equation}
\label{temperature}
T=\frac{\hbar}{8 \pi \left( M + \frac{\alpha_1 \hbar M}{4(\hbar +M^2 )} \right)}~.
\end{equation}
If $\alpha _1$ is independent of $M$ then as $M \rightarrow 0$ one has $T \rightarrow \infty$. By allowing
$\alpha _1$ to depend on $M$ one can have both a finite entropy and temperature as $M \rightarrow 0$. One
example of this is the choice
\begin{equation}
\label{alpha1}
\alpha_1 = K \frac{\hbar + M^2}{M^2 + \sqrt{\hbar} M}
\end{equation}
where $K$ is a dimensionless constant. The particular combination of $M$ and $\hbar$ is required to make
$\alpha _1$ dimensionless. This choice of $\alpha _1$ gives temperature and entropy of the form
\begin{equation}
\label{ts1}
T=\frac{\hbar}{8 \pi \left( M + \frac{K \hbar}{4(\sqrt{\hbar} +M )} \right)}~ ~ ~ ~  , ~ ~ ~ ~
S= \frac{4 \pi M^2}{\hbar} + 2 \pi K \ln \left( 1 + \frac{M}{\sqrt{\hbar}} \right)~.
\end{equation}
This expression for the entropy is essentially the same as in (\ref{entropy-corrected2}) with $\alpha _1 \rightarrow
2 K$ and $M^2/\hbar \rightarrow M/\sqrt{\hbar}$ inside the logarithm. For the temperature given above
we find $T \rightarrow \sqrt{\hbar}/2 \pi K$ as $M \rightarrow 0$. In the following we will simply take $\alpha _1$ to be
independent of $M$ and $\hbar$ since to address the information loss puzzle we only need the entropy to take
the form given in (\ref{entropy-corrected2}) or (\ref{ts1}) which happens whether we pick $\alpha _1$ as in
(\ref{alpha1}) or let $\alpha _1$ be independent of $M$ and $\hbar$. In figures (\ref{entropy}) and
(\ref{temp}) we show the behavior of the entropy and temperature for both positive and negative $\alpha _1$. The
dashed line is the lowest order result; the dotted line is the lowest order plus first order quantum correction;
the dashed line is the all orders results for our particular choice of $\alpha _i$'s.

We now want to show that the original entropy of the black hole given in equation (\ref{entropy-corrected2}) can be
completely carried away by the emitted radiation when specific conditions are satisfied.
Thus, the entropy is conserved which implies a unitary evolution
and thus under specific conditions gives a possible resolution to the information loss paradox. To begin we look at the connection
between  the tunneling rate (\ref{gamma}) and the change in entropy as given in \cite{Wilczek}
\begin{eqnarray}
\label{entropy change2}
\Gamma =e^{\Delta S}
\end{eqnarray}
where $\Delta S=S(M-\omega)-S(M)$ is the entropy change during the tunneling
process when the black hole mass goes from $M$ to $M-\omega$ due to the emission of a field quantum
of energy $\omega$. Using (\ref{entropy-corrected2}) one finds the following
expression for $\Delta S$
\begin{equation}
\label{DeltaS}
\Delta S=-\frac{8\pi}{\hbar}\omega \left( M-\frac{\omega}{2} \right)
+ \pi\alpha_1 \ln \left[ \frac{1+ (M-\omega)^2 / \hbar}{1+  M ^2 / \hbar} \right].
\end{equation}
Now combining equations (\ref{entropy change2}) and (\ref{DeltaS}), the corrected tunneling rate takes the form
\begin{eqnarray}
\label{tunneling}
\Gamma (M; \omega)= \left( \frac{1+ (M-\omega)^2 / \hbar}{1+  M ^2 / \hbar}\right)^{\pi\alpha_1}\exp{\left[-\frac{8\pi}{\hbar}\omega
\left(M-\frac{\omega}{2} \right) \right]} .
\end{eqnarray}
The term $\exp{\left[-\frac{8\pi}{\hbar}\omega
\left(M-\frac{\omega}{2} \right) \right]}$ represents the result of back reaction
on the tunneling rate while the term to the power $\pi \alpha_1$ represents the
quantum corrections to all orders. Even in the classical limit, i.e. setting
$\pi \alpha_1 =0$, there is a deviation from a thermal spectrum
due to the $\omega ^2$ term in the exponent.

We now find the connection between the tunneling rate (\ref{tunneling}) and the entropy of
the emitted radiation, $S_{rad}$. Assuming that the black hole mass is completely radiated away
we have the relationship $M=\omega_1 + \omega_2 +...+ \omega _n = \sum _{j=1} ^n \omega _j$ between the mass of the black hole
and the energy, $\omega_j$, of the emitted field quanta. The probability for this radiation to occur is
given by the following product of $\Gamma$'s which are defined in equation (\ref{tunneling}) \cite{information2}
\begin{equation}
\label{probability}
P _{rad} = \Gamma (M; \omega_1) \times \Gamma (M-\omega_1 ; \omega _2) \times ... \times
\Gamma \left( M- \sum _{j=1} ^{n-1} \omega _j ; \omega _n \right)
\end{equation}
where the probability of emission of the individual field quanta of energy $\omega _j$ is given by
\begin{eqnarray}
\label{probability2}
\Gamma (M; \omega_1) &=& \left( \frac{1+ (M-\omega_1)^2 / \hbar}{1+ M ^2 / \hbar}\right)^{\pi\alpha_1}\exp{\left[-\frac{8\pi}{\hbar}\omega_1
\left(M-\frac{\omega_1}{2} \right) \right]} ~, \nonumber \\
\Gamma (M-\omega_1; \omega_2) &=& \left( \frac{1+  (M-\omega_1 -\omega_2)^2 / \hbar}{1+  (M-\omega_1) ^2 / \hbar}\right)^{\pi\alpha_1}
\exp{\left[-\frac{8\pi}{\hbar}\omega_2  \left(M -\omega_1 - \frac{\omega_2}{2} \right) \right]} ~, \nonumber \\
&\,& \nonumber \\
&&..... ~, \\
&\,& \nonumber \\
\Gamma \left( M- \sum _{j=1} ^ {n-1} \omega_j; \omega_n \right) &=&
\left( \frac{1+ (M- \sum _{j=1} ^ {n-1} \omega_j -\omega_n)^2 / \hbar}{1+  (M-\sum _{j=1} ^ {n-1} \omega_j ) ^2 / \hbar}\right)^{\pi\alpha_1}
\exp{\left[-\frac{8\pi}{\hbar}\omega_n  \left(M - \sum _{j=1} ^ {n-1} \omega_j - \frac{\omega_n}{2} \right) \right]} \nonumber \\
&=& \left( \frac{1}{1+ (M-\sum _{j=1} ^ {n-1} \omega_j ) ^2 / \hbar}\right)^{\pi\alpha_1} \exp( -4 \pi  \omega _n ^2 / \hbar) ~. \nonumber
\end{eqnarray}
The $\Gamma$'s of the form $\Gamma (M-\omega_1 -\omega_2-...-\omega_{j-1} ; \omega_j)$ represent the probability for the
emission of a field quantum of energy $\omega _j$ with the condition that first field quanta of energy $\omega_1 + \omega_2+...+\omega_{j-1}$
have been emitted.

Now using (\ref{probability2}) in equation (\ref{probability}) we find the total probability for the radiation process described above
\begin{equation}
\label{probability3}
P_{rad} = \left( \frac{1}{1+ M ^2 / \hbar}\right)^{\pi\alpha_1} \exp( -  4\pi M ^2 / \hbar) ~.
\end{equation}
The black hole mass could also have been radiated away by a different sequence of field quanta energies e.g.
first $\omega_2$ is emitted and then other field quanta are emitted in the following order $\omega_2 +\omega_1+...+\omega_{n-1} + \omega_n$.
Assuming each of these different processes has the same probability one can count the number of
microstates, $\Omega$, for the above process as $\Omega = 1/P_{rad}$. Then using the
Boltzmann definition of entropy as the natural logarithm of the number of microstates we have
\begin{equation}
\label{rad-entropy}
S_{rad} = \ln (\Omega ) = \ln ( 1/ P_{rad}) = \frac{4 \pi}{\hbar} M^2 + \pi \alpha _1 \ln \left( 1 + \frac{M^2}{\hbar} \right) ~.
\end{equation}
This entropy of the emitted radiation (\ref{rad-entropy}) is identical to the original entropy of the black hole (\ref{entropy-corrected2}), thus
entropy is conserved between the initial (black hole plus no radiation) and final (no black hole plus radiated field quanta) states.
This implies the same number of microstates between the initial and final states and thus unitary evolution.
This then provides a possible resolution of the information paradox when the specific conditions are imposed.

\section{Discussion and Conclusion}

In this article we have used the WKB/tunneling method of calculating Hawking radiation to obtain several
non-trivial results concerning the evaporation of black holes and the information paradox taking into
account both quantum corrections to all orders in $\hbar$ and/or back reaction. In section II we showed that
quantum corrections to all orders in $\hbar$, as parameterized by the unknown $\alpha _j$'s, did not alter the
spectrum -- the spectrum is thermal to all orders in $\hbar$. Thus, quantum corrections alone can not solve
the information loss paradox.

In section III we showed that by taking back reaction and quantum corrections to all orders in $\hbar$ into
account {\it and} assuming some specific form for the quantum corrections (see equation (\ref{alpha})), one could
present a possible resolution for the information loss paradox. In this resolution the black hole completely evaporates, but the entropy
is conserved. The initial quantum corrected formula for the entropy given in equation (\ref{entropy-corrected2}) is equal
to the entropy of the emitted radiation (\ref{rad-entropy}). This conservation of entropy indicates that the initial and
final states have the same number of microstates and thus that the evolution is unitary.

As a final comment we note that our pragmatic choice of $\alpha_j$'s in \eqref{alpha} also
led to well behaved temperature versus mass behavior for the black hole such as given in figure \ref{ta} by the all order
quantum corrections curve (i.e. the solid curve). ``Well behaved" here means that, in contrast to the lowest order Hawking result
(i.e. the dashed curve in figure \ref{ta}) where the temperature diverges as the mass goes to zero, the all order quantum corrections,
with our choice of $\alpha_j$'s, gave a curve which for large mass matched the Hawking result but at some mass reached a maximum
and thereafter decreased to zero as mass went to zero.  This ``well behaved" behavior is similar to that found using the
microcanonical description of Hawking radiation \cite{harms} applied to simple black holes or also the lowest order Hawking radiation
applied non-commutative black holes \cite{nicolini}. Further in the microcanonical description of reference \cite{harms} the total
energy (black hole plus radiation) is conserved and the black completely evaporates in a finite time. Similarly, in our analysis of section III
we found a conservation of entropy (initial black hole entropy equals the entropy of the emitted radiation). Also the black hole
was able to completely evaporate away with the entropy of the initial black hole being converted to entropy of the outgoing radiation.
We hope to explore these similarities in the future with an eye toward finding a firmer motivation for our choice of $\alpha_j$'s in \eqref{alpha}.

\begin{acknowledgments}
This work was supported by the National Natural Science Foundation
of China (No.10275030) and the Cuiying Programme of Lanzhou
University (225000-582404). The authors would like to thank Dr. Izzet Sakalli for
pointing out the mistake in the expression for the temperature in the original article.
 \end{acknowledgments}

\end{document}